\documentclass[UTF-8,6pt,a4paper]{article}
\usepackage{bbm}
\usepackage{amsfonts}
\usepackage{mathrsfs}
\usepackage {amssymb}
\usepackage {amsmath}
\usepackage{amsthm}
\usepackage{latexsym}
\def\dse#1{\vskip 0.6cm\noindent
        {\large\bf #1}
        \vskip 0.4cm}

\def\dse#1{\vskip 0.6cm\noindent
        {\large\bf #1}
        \vskip 0.4cm}
\usepackage{lastpage}
\usepackage{epsfig}

\makeatletter
    
    \newcommand{\Rmnum}[1]{\expandafter\@slowromancap\romannumeral #1@}
  \makeatother

\begin{document}
\begin{center}
\textbf{\large{New  quantum codes from
 constacyclic codes over finite chain rings}}
\footnote { E-mail addresses: ysh$_{-}$tang@163.com(Y.Tang),
yaoting$_{-}$1649@163.com(T.Yao),
 693204301@qq.com(H.Xu),\\  kxs$6$@sina.com(X.Kai).\\
$^*$This research is supported by National Natural Science Funds of
China (Nos. 12171134, 62172183 and 12201170), Natural Science
Foundation of Anhui Province (No. 2108085QA03), Key University
Science Research Project of  Anhui Province (No. KJ2021A0926).}
\end{center}

\begin{center}
{Yongsheng Tang$^{1}$, Ting Yao$^{1}$, Heqian Xu$^{1}$, Xiaoshan
Kai$^{2}$ }
\end{center}

\begin{center}
\textit{1.School of Mathematics and Statistics, Hefei Normal
University, Hefei, 230601, Anhui, P.R.China;\\ 2. School of
Mathematics, Hefei University of Technology, Hefei, 230009, Anhui,
P.R.China }
\end{center}

\noindent\textbf{Abstract} \ Let $R$ be the finite chain ring
$\mathbb{F}_{p^{2m}}+{u}\mathbb{F}_{p^{2m}}$, where
$\mathbb{F}_{p^{2m}}$ is the finite field with $p^{2m}$ elements,
 $p$ is a prime, $m$ is  a non-negative integer and
${u}^{2}=0.$  In this paper, we firstly define a class of Gray maps,
which changes the Hermitian self-orthogonal property of linear codes
over $\mathbb{F}_{2^{2m}}+{u}\mathbb{F}_{2^{2m}}$ into the Hermitian
self-orthogonal property of linear codes over $\mathbb{F}_{2^{2m}}$.
Applying the Hermitian construction,   a new class of $2^{m}$-ary
 quantum codes are obtained from
Hermitian  constacyclic self-orthogonal codes over
$\mathbb{F}_{2^{2m}}+{u}\mathbb{F}_{2^{2m}}.$  We secondly  define
another class of  maps, which changes the Hermitian self-orthogonal
property of linear codes over $R$ into the  trace self-orthogonal
property of linear codes over $\mathbb{F}_{p^{2m}}$. Using the
Symplectic construction,   a new class of $p^{m}$-ary quantum codes
are obtained from Hermitian constacyclic self-orthogonal codes over
$R.$

\noindent\emph{keywords}: quantum  code; Symplectic construction;
Hermitian construction; constacyclic code

\section{Introduction}
\label{sec:1} Quantum error-correcting codes provide detection or
correction of the errors which occur in a noisy quantum
communication channel. In [8], it is shown that it is equivalent
finding additive codes over $\mathbb{F}_{4}$ which are
self-orthogonal with respect to certain trace inner product to
finding binary stabilizer quantum error-correcting codes. The
results given in [8] are generalized to stabilizer quantum
error-correcting codes over $\mathbb{F}_{q}$ in [21]. Consequently,
constructions of  quantum error-correcting codes with good
parameters have been developed rapidly. Many classes of quantum
codes with good parameters were constructed from cyclic codes or
constacyclic codes over finite fields by using
Calderbank-Shor-Steane (CSS) construction, Hermitian construction
and Symplectic construction, respectively (see [1,9,19,20,22]).

Recently, many researchers have applied cyclic codes or constacyclic
codes over finite rings to construct quantum codes with good
parameters.  In these studies, the finite rings divided into two
classes in general. One is  finite chain rings, the other is  finite
non-chain rings. We firstly introduce linear codes over finite chain
rings have been used to construct  quantum codes with good
parameters.   Quantum codes with good parameters have been
constructed from cyclic codes over $\mathbb{F}_{2}+u\mathbb{F}_{2}$
with  $u^{2}=0$ in [29]. Later, this construction method is
generalized to cyclic codes over $\mathbb{F}_{4}+u\mathbb{F}_{4}$
with $u^{2}=0$ in [18], and
$\mathbb{F}_{2^{m}}+u\mathbb{F}_{2^{m}}+\cdots+u^{k}\mathbb{F}_{2^{m}}
$ with $u^{k+1}=0$ in [32]. Liu et al.  obtained  quantum codes with
good parameters from linear  codes over
$\mathbb{F}_{p^{2m}}+u\mathbb{F}_{p^{2m}}$ with $u^{2}=0$ in [25].
Ding et al.  obtained   quantum codes with good parameters from
 constacyclic codes over
$\mathbb{F}_{p^{m}}+u\mathbb{F}_{p^{m}}$ with $u^{2}=0$ in [12].
Tang et al. obtained  quantum codes with good parameters from
constacyclic codes over $\mathbb{F}_{2^{m}}+u\mathbb{F}_{2^{m}}$
with $u^{2}=0$ in [33] and [34], respectively.

We secondly introduce linear codes over finite non-chain  rings have
been used to construct  quantum codes with good parameters.  Qian
obtained quantum codes with good parameters from cyclic codes over
$\mathbb{F}_{2}+v\mathbb{F}_{2}$ with $v^{2}=v$ in [30]. Ashraf and
Mohammad obtained quantum codes with good parameters from cyclic
codes over $\mathbb{F}_{3}+v\mathbb{F}_{3}$ with $v^{2}=1$ in [2],
and $\mathbb{F}_{p}+v\mathbb{F}_{p}$ with $v^{2}=v$ in [3],
respectively. Sari and Siap obtained  quantum codes with good
parameters from cyclic codes over
$\mathbb{F}_{p^{r}}+v\mathbb{F}_{p^{r}}+\cdots+v^{m-1}\mathbb{F}_{p^{r}}$
with $v^{m}=v$ in [31].  Gao et al. obtained quantum codes with good
parameters from cyclic codes over
$\mathbb{F}_{q}+v_{1}\mathbb{F}_{q}+\cdots+v_{r}\mathbb{F}_{q}$ with
$v_{i}^{2}=v_{i},v_{i}v_{j}=v_{j}v_{i}=0$ in [15]. Dertli et al.
obtained  quantum codes with good parameters from cyclic codes over
$\mathbb{F}_{2}+u\mathbb{F}_{2}+v\mathbb{F}_{2}+uv\mathbb{F}_{2}$,
where $u^{2}=u,v^{2}=v,uv=vu$ in [10]. Ashraf and Mohammad obtained
  quantum codes with good
parameters from cyclic codes over
$\mathbb{F}_{q}+u\mathbb{F}_{q}+v\mathbb{F}_{q}+uv\mathbb{F}_{q}$
with $u^{2}=u,v^{2}=v,uv=vu$ in [4], and
$\mathbb{F}_{p}[u,v]/\langle u^{2}-1,v^{3}-v,uv-vu\rangle$ with
$u^{2}=1,v^{3}=v,uv=vu$ in [5], respectively.  Bag et al. obtained
 quantum codes with good
parameters from constacyclic codes over
$\mathbb{F}_{p^{m}}+v\mathbb{F}_{p^{m}}$ with $v^{2}=1,$ $p$ is odd
prime in [6]. Gao et al. obtained quantum codes with good parameters
from constacyclic codes over $\mathbb{F}_{p}+u\mathbb{F}_{p}$ with
$u^{2}=1$ in [14]. Wang et al. obtained   quantum codes with good
parameters from constacyclic codes over
$\mathbb{F}_{p}+v\mathbb{F}_{p}$ with $v^{2}=v$ in [36]. Ma et al.
obtained    quantum codes with good parameters from constacyclic
codes over $\mathbb{F}_{q}+v\mathbb{F}_{q}+v^{2}\mathbb{F}_{q}$ with
$v^{3}=v$ in [26]. They  also obtained  quantum codes with good
parameters from constacyclic codes over $\mathbb{F}_{q}[u,v]/\langle
u^{2}-1,v^{2}-v,uv-vu\rangle$ with $u^{2}=1,v^{2}=v,uv=vu$ in [27].
Bag et al. obtained  quantum codes with good parameters from a class
of constacyclic codes over $\mathbb{F}_{p}[u]/\langle u^{k+1}-u
\rangle$ with  $p$ is odd prime, $k$ is an integer and
$\textrm{gcd}(p,k)=1$ in [11], and
$\mathbb{F}_{q^{2}}+u\mathbb{F}_{q^{2}}$ with $u^{2}=\omega^{2}$ in
[7], respectively. Wang et al. obtained  quantum codes with good
parameters from constacyclic codes over
$\mathbb{F}_{q^{2}}+v\mathbb{F}_{q^{2}}$ with $v^{2}=v$ in [37], and
$\mathbb{F}_{q^{2}}+v_{1}\mathbb{F}_{q^{2}}+\cdots+v_{r}\mathbb{F}_{q^{2}}$
with $v_{i}^{2}=v_{i},v_{i}v_{j}=v_{j}v_{i}=0$ in [38],
respectively. Ji and Zhang obtained  quantum codes with good
parameters from constacyclic codes over
$\mathbb{F}_{q}[u_{1},u_{2},\cdots,u_{k}]/\langle
u_{i}^{3}-u_{i},u_{i}u_{j}-u_{j}u_{i}\rangle$ in [16].

Motivated by these excellent works, we consider two constructions of
quantum codes from
 Hermitian  $\alpha(1+u)$-constacyclic self-orthogonal codes over the finite chain  ring
$\mathbb{F}_{p^{2m}}+{u}\mathbb{F}_{p^{2m}}$ with $u^{2}=0$. By
applying the Hermitian construction and the Symplectic construction,
respectively,  we obtain a new class of $2^{m}$-ary quantum codes
and a new class of $p^{m}$-ary quantum codes from Hermitian
 $\alpha(1+u)$-constacyclic self-orthogonal codes over
$\mathbb{F}_{2^{2m}}+u\mathbb{F}_{2^{2m}}$ and
$\mathbb{F}_{p^{2m}}+u\mathbb{F}_{p^{2m}},$ respectively.  Compared
our obtained quantum codes with the existing codes available in
Ref.[13], we find that the parameters of our obtained quantum codes
are optimal   at present. Moreover, we find that some our obtained
quantum codes are satisfying $2d=n-k.$

The rest of this paper is organized as follows. In Section 2, we
introduce some basic definitions and notations of linear codes over
$\mathbb{F}_{p^{2m}}+{u}\mathbb{F}_{p^{2m}}$. A class of Gray maps
$\Phi_{M}$ is defined in Section 3. In Section 4, we construct a new
class of  $2^{m}$-ary quantum codes based on Hermitian
 $\alpha(1+u)$-constacyclic self-orthogonal codes over
$\mathbb{F}_{2^{2m}}+u\mathbb{F}_{2^{2m}}$  by using the Hermitian
construction. Some optimal $p^{m}$-ary quantum codes from Hermitian
 $\alpha(1+u)$-constacyclic self-orthogonal codes over
$\mathbb{F}_{p^{2m}}+u\mathbb{F}_{p^{2m}}$  via  the Symplectic
construction are obtained in Section 5. Section 6 concludes the
paper.

\section{Preliminaries}
\label{sec:1} Let $\mathbb{F}_{p^{2m}}$ be the finite field with
$p^{2m}$ elements. Let  $R$ be the polynomial residue  ring
$\mathbb{F}_{p^{2m}}+u\mathbb{F}_{p^{2m}}$ with $u^2=0$. The ring
$R$ is a finite chain ring with a maximal ideal $\langle u\rangle$.
The units of $R$ are the elements $\{a+ub\mid a \neq 0\}$ and the
residue field is $R/uR\simeq\mathbb{F}_{p^{2m}}$. A code of length
$N$ over $R$ is a nonempty subset of $R^N$, and a code is linear
over $R$ if it is an $R$-submodule of $R^N$. Two codes are
 permutation equivalent if one can be obtained from the other by permuting the
coordinates. Any code over $R$ is permutation equivalent to a code
$C$ with generator matrix of the form:
$$G=
\left
(\begin {array} {ccc} I_{k_{0}} & A & B \\
0 & uI_{k_{1}} & uD
\end {array}\right),$$ where  $B$ is  a matrix   over $R$ and $A, \ D$ are
$\mathbb{F}_{p^{2m}}$ matrices. Then $C$ is an abelian group of type
$\{k_{0},k_{1}\}$, $C$ contains $p^{4m(2k_{0}+k_{1})} $ codewords,
and $C$ is a free $R$-module if and only if $k_{1}=0$.

For any element $c=a+ub\in R,$  where $a,b \in \mathbb{F}_{p^{2m}}$.
The conjugation of $c=a+ub$ is defined by
$\overline{c}=\overline{a}-u\overline{b},$   where
$\overline{a}=a^{p^{m}},\overline{b}=b^{p^{m}}.$ Naturally, the
conjugation of $\mathbf{c}=(c_{0}, c_{1}, \cdots, c_{N-1})\in R^{N}$
is defined as $\overline{\mathbf{c}}=(\bar{c}_{0}, \bar{c}_{1},
\cdots, \bar{c}_{N-1})$. For any $\mathbf{x}=(x_{0}, x_{1}, \cdots,
x_{N-1})$ and $\mathbf{y}=(y_{0}, y_{1}, \cdots, y_{N-1})$, the
Hermitian inner product of $\mathbf{x}$ and $\mathbf{y}$ is defined
as

\begin{center}
$\langle\mathbf{x}, \mathbf{y}\rangle_{H}
=x_{0}\bar{y}_{0}+x_{1}\bar{y}_{1}+\cdots+x_{N-1}\bar{y}_{N-1} $.
\end{center}

\noindent The Hermitian dual code $C^{\bot_{H}}$ of $C$ is define as

\begin{center}
$C^{\bot_{H}}=\{\mathbf{x}\in R^{N}|\ \langle\mathbf{x},
\mathbf{y}\rangle_{H}=0$, for all $\mathbf{y}\in C\}$.
\end{center}

\noindent A code $C$ is called Hermitian self-orthogonal if $C
\subseteq C^{\bot_{H}}$, and Hermitian self-dual if $C^{\bot_{H}} =
C$.

Similar to Proposition 1.2 in Ref.[35], we easily obtain  the
generator matrix of the Hermitian dual code of $C$.\\

\noindent\textbf{Proposition 2.1.} \emph{The Hermitian dual code
$C^{\perp H}$ of the linear code $C$ of length $N$ over $R$ with
generator matrix above has generator matrix}
\begin{displaymath}
\left(\begin{array}{ccc}
\overline{B}^{T}+\overline{D}^{T} \ \overline{A}^{T}&\overline{D}^{T}&I_{N-k_{0}-k_{1}}\\
u\overline{A}^{T}&uI_{k_{1}}&0\\
\end{array}\right),
\end{displaymath}
\emph{where $\overline{A}^{T}, \overline{B}^{T}$ and
$\overline{D}^{T}$ denote the conjugate transposes of $A,B$ and $D$,
respectively. Moreover, $C^{\perp H}$ is an abelian group of type
$\{N-k_{0}-k_{1},k_{1}\}$ , and its size is
$p^{2m(2N-2{k_{0}}-{k_{1}})}$.}\\

\noindent Let $C$ be a code of length $N$ over $R$ and $P(C)$ be its
polynomial representation, i.e.,
$$P(C)=\left\{\sum_{i=0}^{N-1}c_ix^i\mid(c_0,c_1,\ldots,c_{N-1})\in C\right\}.$$

Let $\sigma_{\lambda}$ be a map from $R^N$ to $R^N$ given by
$$\sigma_{\lambda}(c_0,c_1,\ldots, c_{N-1})=(\lambda c_{N-1}, c_0,\ldots,c_{N-2}),$$
where $\lambda \in R$. Then $C$ is said to be cyclic if
$\sigma_1(C)=C$ and $\alpha(1+u)$-constacyclic if
$\sigma_{\alpha(1+u)}(C)=C$. A code $C$ of length $N$ over $R$ is
cyclic if and only if $P(C)$ is an ideal of $R[x] / {\langle
x^N-1\rangle}$ and a code $C$ of length $N$ over $R$ is
$\alpha(1+u)$-constacyclic if and only if $P(C)$ is an ideal of
$R[x]/ {\langle x^N-\alpha(1+u)\rangle}$. Throughout this paper, we
assume that
$\alpha(1+u)$-constacyclic codes over $R$ of length $N=p^{e}n,$  where $e$ is a non-negative integer and $n$  and $p$ are coprime.\\

\section{A class of Gray maps over $R$}

 A
class of Gray  maps $\Phi_{M}$ from
$R=\mathbb{F}_{p^{2m}}+{u}\mathbb{F}_{p^{2m}}$ to
$\mathbb{F}_{p^{2m}}^{2}$ is defined as follows.

\noindent\textbf {Definition 3.1}  A  map $\Phi_{M}$ on $R$ is
defined as
\begin{center}
$\Phi_{M}: R \rightarrow\mathbb{F}_{p^{2m}}^{2}$,\\
 $r
+uq \mapsto(q,  r+q)M,$
\end{center}
\noindent where $ M=$ $\begin{pmatrix}
a\  &  b  \\
s\  &  t  \\
\end{pmatrix}
\in GL_{2}(\mathbb{F}_{p^{2m}}),$ and $GL_{2}(\mathbb{F}_{p^{2m}})$ is the set of all $2 \times 2$ invertible matrices over $\mathbb{F}_{p^{2m}}$.\\
The
 map $\Phi_{M}$ can be naturally extended to $R^{n}$ as follows:
\begin{center}
$\Phi_{M}:R^{N}\rightarrow\mathbb{F}_{p^{2m}}^{2N}$,

$(c_{0}, c_{1}, \ldots, c_{N-1}) \mapsto
(aq_{0}+s(r_{0}+q_{0}),\ldots,aq_{N-1}+s(r_{N-1}+q_{N-1}),
bq_{0}+t(r_{0}+q_{0}),\ldots,bq_{N-1}+t(r_{N-1}+q_{N-1}))$,\\
\end{center}
\noindent where $c_{i}=r_{i}+uq_{i},$ for $0\leq i \leq N-1$.\\

The Hamming weight of $\textbf{c}\in C$, denoted by
$w_{H}(\textbf{c})$, is the number of its nonzero components. The
Hamming distance $d_{H}(\textbf{c}, \textbf{c}')$ between two
codewords $\textbf{c}$ and $\textbf{c}'$ is the  Hamming weight of
the codeword $\textbf{c}-\textbf{c}'$. The Hamming distance $d_H(C)$
of $C$ is defined as $\min\{w_H(\textbf{c}-
\textbf{c}')|\textbf{c},\textbf{c}'\in C, \textbf{c}\neq
\textbf{c}'\}$. That is, for any linear code, the Hamming distance
$d_H(C)$
of $C$ is its minimum Hamming weight.\\

 The Gray weight of any
element $r+uq \in R$ is defined as the Hamming weight of
$\Phi_{M}(r+uq)$, i.e., the number of nonzero components of
$\Phi_{M}(r+uq)$. The Gray weight of a codeword $\mathbf{c}=(c_{0},
c_{1}, \cdots, c_{N-1})\in R^{N}$ is defined to be the rational sum
of the Gray weight of its components. For any
$\mathbf{x},\mathbf{y}\in R^{N}$,  the Gray distance is given by
$d_{G}(\mathbf{x},\mathbf{y})=w_{G}(\mathbf{x}-\mathbf{y})$. The
 Gray distance $d_{G}(C)$ of a code $C$ is the smallest
nonzero Gray distance between all pairs of distinct codewords. The
minimum Gray weight $w_{G}(C)$ of $C$ is the smallest nonzero Gray
weight among all codewords. If $C$ is linear, then the Gray
distance $d_{G}(C)$ is the same as the minimum Gray weight $w_{G}(C)$.\\

\noindent\textbf{Proposition 3.2.} \emph{The Gray map $\Phi_{M}$ is
a distance-preserving map
from ($R^N$, Gray distance) to ($\mathbb{F}_{p^{2m}}^{2N}$, Hamming distance) and it is also $\mathbb{F}_{p^{2m}}$-linear.}\\

\noindent{\it Proof} \ \ Let $k_{1},k_{2}\in\mathbb{F}_{p^{2m}}$,
$\mathbf{x},\mathbf{y}\in R^{N}$, from the definition of Gray map,
we have
$\Phi_{M}(k_{1}\mathbf{x}+k_{2}\mathbf{y})=k_{1}\Phi_{M}(\mathbf{x})
+k_{2}\Phi_{M}(\mathbf{y})$, which means that $\Phi_{M}$ is an
$\mathbb{F}_{p^{2m}}$-linear map. Moreover,
\begin{align*}
d_{G}(\mathbf{x},\mathbf{y})&=w_{G}(\mathbf{x}-\mathbf{y})\\
&=w_{H}(\Phi_{M}(\mathbf{x}-\mathbf{y}))\\
&=w_{H}(\Phi_{M}(\mathbf{x})-\Phi_{M}(\mathbf{y}))\\
&=d_{H}(\Phi_{M}(\mathbf{x}),\Phi_{M}(\mathbf{y})).
\end{align*}
This completes the proof. \qed \\

\noindent\textbf {Lemma 3.3.} \emph{Let $ M=$ $\begin{pmatrix}
a\  &  b  \\
s\  &  t  \\
\end{pmatrix}
\in GL_{2}(\mathbb{F}_{p^{2m}})$,  the conjugation matrix of $ M$ be
$ \overline{M}=$ $\begin{pmatrix}
\bar{a}\  &  \bar{b}  \\
\bar{s}\  &  \bar{t}  \\
\end{pmatrix}.$  Then
\begin{center}
$ M\overline{M}^{T}=\Omega=\begin{pmatrix}
\lambda& 0  \\
0 & \lambda  \\
\end{pmatrix}$
$(\lambda\in\mathbb{F}_{p^{2m}}^{*}=\mathbb{F}_{p^{2m}}\backslash\{0\})$
\end{center}}

\noindent \emph{if and only if $a\bar{s}+b\bar{t}=0$ and $a\bar{a}+b\bar{b}=s\bar{s}+t\bar{t}=\lambda \neq 0$.}\\

\noindent{\it Proof} \ \  Note that
$M\overline{M}^{T}=\begin{pmatrix}
a\bar{a}+b\bar{b} & a\bar{s}+b\bar{t}  \\
\bar{a}s+\bar{b} t & s\bar{s}+t\bar{t}  \\
\end{pmatrix}$.
The desired result follows. \qed\\

\noindent\textbf{Theorem 3.4.} \emph{Let $C$ be a linear code of
length $N'=2^{e}n$ over $\mathbb{F}_{2^{2m}}+u\mathbb{F}_{2^{2m}},$
where $e$ is a non-negative integer and $n$ is odd.  Let $ M=$
$\begin{pmatrix}
a\ & b \\
s\ & t \\
\end{pmatrix}\in GL_{2}(\mathbb{F}_{2^{2m}})$
and $M\overline{M}^{T}=\Omega$. Then
$\Phi_M(C^{\bot_{H}})=\Phi_M(C)^{\bot_{H}}$. Moreover, if
$C^{\bot_{H}} \subseteq C$, then $\Phi_M(C^{\bot_{H}})\subseteq
\Phi_M(C)^{\bot_{H}}$.}\\

\noindent{\it Proof} \ \  Let
$\textbf{c}_{1}=\textbf{r}_{1}+u\textbf{q}_{1}\in C$ and
$\textbf{c}_{2}=\textbf{r}_{2}+u\textbf{q}_{2}\in C^{\bot_{H}}$,
where $\mathbf{r}_i=(r_{i1},r_{i2},\cdots, r_{i N'})$ and
$\mathbf{q}_i=(q_{i1},q_{i2},\cdots, q_{i N'})\in
\mathbb{F}_{2^{2m}}^{N'}$. The Hermitian inner product of
$\textbf{c}_{1}$ and $\textbf{c}_{2}$ is
\begin{align*} \langle
\textbf{c}_{1},\textbf{c}_{2}\rangle_{H}
&=\sum_{i=1}^{N'} (r_{1i}+u q_{1i})(\overline{r_{2i}}+u\overline{q_{2i}})\\
&=\sum_{i=1}^{N'} r_{1i}\overline{r_{2i}}+u \sum_{i=1}^{N'} (r_{1i}\overline{q_{2i}}+\overline{r_{2i}}q_{1i}) \\
&=\langle\textbf{r}_{1},\textbf{r}_{2}\rangle_H
+u(\langle\textbf{r}_{1},\textbf{q}_{2}\rangle_H+
\langle\textbf{q}_{1},\textbf{r}_{2}\rangle_H).
\end{align*}
It follows from $\langle \textbf{c}_{1},\textbf{c}_{2}\rangle_{H}=0$
that $\langle\textbf{r}_{1},\textbf{r}_{2}\rangle_H=0$ and
$\langle\textbf{r}_{1},\textbf{q}_{2}\rangle_H+
\langle\textbf{q}_{1},\textbf{r}_{2}\rangle_H=0$.

\noindent On the other hand, \begin{align*}
&\langle\Phi_{M}(\textbf{c}_{1}),\Phi_{M}(\textbf{c}_{2})\rangle_{H}\\
=~&\sum_{i=1}^{N'} [a q_{1i}+s(r_{1i}+q_{1i})] \overline{[a q_{2i}+s(r_{2i}+q_{2i})]}+\sum_{i=1}^{N'} [b q_{1i}+t(r_{1i}+q_{1i})] \overline{[b q_{2i}+t(r_{2i}+q_{2i})]}\\
=~& (a\overline{a}+b\overline{b})  \langle\textbf{q}_{1},\textbf{q}_{2}\rangle_H + (a\overline{s}+b\overline{t})(\langle\textbf{q}_{1},\textbf{r}_{2}\rangle_H+\langle\textbf{q}_{1},\textbf{q}_{2}\rangle_H)+(s\overline{a}+t\overline{b})(\langle\textbf{r}_{1},\textbf{q}_{2}\rangle_H+\langle\textbf{q}_{1},\textbf{q}_{2}\rangle_H)\\
&+(s\overline{s}+t\overline{t})(\langle\textbf{r}_{1},\textbf{r}_{2}\rangle_H+\langle\textbf{r}_{1},\textbf{q}_{2}\rangle_H+\langle\textbf{q}_{1},\textbf{r}_{2}\rangle_H+\langle\textbf{q}_{1},\textbf{q}_{2}\rangle_H)\\
=~&\lambda[ \langle\textbf{r}_{1},\textbf{r}_{2}\rangle_H+\langle\textbf{r}_{1},\textbf{q}_{2}\rangle_H+\langle\textbf{q}_{1},\textbf{r}_{2}\rangle_H ],\\
=~&0,
\end{align*}
where the third equation follows from Lemma 3.3. Therefore,
$\Phi_{M}(C^{\bot_{H}})\subseteq \Phi_{M}(C)^{\bot_{H}}$.

\noindent Suppose $|C|=2^{4m(2k_0+k_1)}$. Since $\Phi_{M}$ is a
bijection,
\begin{align*}
|\Phi_{M}(C^{\bot_{H}})|    &=|C^{\bot_H}|\\
&=2^{2m(2N'-2k_0-k_1)}\\
&=2^{4mN'}/|\Phi_M(C)|\\
&=|\Phi_M(C)^\perp|.
\end{align*}
Therefore, $\Phi_{M}(C^{\bot_{H}})= \Phi_{M}(C)^{\bot_{H}}$. The
second desired result directly follows. \qed

\section{Hermitian
 construction}

In this section, we obtain a new family of $2^{m}$-ary quantum codes
by using Hermitian
  $\alpha(1+u)$-constacyclic  self-orthogonal codes over $\mathbb{F}_{2^{2m}}+u\mathbb{F}_{2^{2m}}$  of length $N'=2^{e}n.$    We firstly introduce some  definitions and notations of
 polynomials in
$R[x].$
\\

Let $f(x)=a_{k}x^{k}+a_{k-1}x^{k-1}+\cdots+a_{0}$ be a polynomial in
$R[x],$  where $a_{0}$ is an invertible element in $R.$ The
conjugation polynomial of $f(x)$ is defined by
$\overline{f(x)}=\overline{a_{k}}x^{k}+\overline{a_{k-1}}x^{k-1}+\cdots+\overline{a_{0}}$.
 The reciprocal polynomial of $f(x)$ is defined by $f^{*}(x)=x^{k}f(x^{-1})$, i.e., $f^{*}(x)=a_{0}x^{k}+a_{1}x^{k-1}+\cdots+a_{k}$.  Obviously,
$(f^{*}(x))^{*}=f(x)$ and $(f(x)g(x))^{*}=f^{*}(x)g^{*}(x)$. We
denote by $f^{\dagger}(x)$ the conjugation of the reciprocal
polynomial of $f(x)$, i.e. $f^{\dagger}(x)=\overline{f^{*}(x)}.$\\

To study the Hermitian dual of $\alpha(1+u)$-constacyclic  codes, we
need the
following result. Its proof is similar to that of Theorem 4.1 in Ref.[32] and is omitted.\\

\noindent\textbf{Theorem 4.1.}  \emph{Let $C$ be an
$\alpha(1+u)$-constacyclic code over $R$ of length $N=p^{e}n$ ($n$
prime to $p$). Then $C^{\bot H}$ is an
$\overline{\alpha^{-1}}(1+u)$-constacyclic code over $R$ of length
 $N$.}\\

Since $\gcd(p^e, p^{2m}-1)=1$, there exists $\beta \in
\mathbb{F}_{p^{2m}}^*$ such that $\beta^{p^e}=\alpha$. In
$R[x]/\langle x^{N}-\alpha(1+u)\rangle$,
 \begin{align*}
u&~=\alpha^{-1}(x^N+\alpha)\\
&~=\alpha^{-1}(x^n+\beta)^{p^e}.
 \end{align*}
 Therefore, $u\in \langle (x^n+\beta)^{p^e} \rangle$. Let
 $$x^n-\beta= \prod_{i=1}^r f_i(x) $$
 be the factorization of $x^n-\beta$ into irreducible factors over $\mathbb{F}_{p^{2m}}$.  Moreover, we know that  $\alpha(1+u)$-constacyclic  codes over $R$  of length
 $N$ are precisely ideals of $R[x]/\langle x^{N}-\alpha(1+u)\rangle $ and the equation  $(x^{n}-\xi(1+u))^{2L}=(x^{n}-\xi)^{2L}$  always  holds in $R[x]$,  where $\xi \in \mathbb{F}_{p^{2m}}$ and $L$ is  any integer.
  Similar to Theorem 3.4 of Ref.[17], we get the following result.\\

\noindent\textbf{Theorem 4.2.} \emph{Let
$x^{n}-\beta=\prod_{i=1}^{r}f_{i}(x)$ be the unique factorization of
$x^{n}-\beta$ into a product of monic basic irreducible pairwise
coprime polynomials in $\mathbb{F}_{p^{2m}}[x],$ where
$\emph{gcd}(n,p)=1$. If $C$ is an $\alpha(1+u)$-constacyclic code of
length $N=p^{e}n$ over $R$, then
$C=\langle\prod_{i=1}^{r}f_{i}^{k_{i}}(x)\rangle$, where $0\leq
k_{i}\leq p^{e+1}$ and
$|C|=p^{2m(2N-\sum_{i=1}^{r}k_{i}\emph{deg}(f_{i}(x)))}$.}\\

Now, we give the structure of the Hermitian dual of $\alpha(1+u)$-constacyclic  codes of  length $N=p^{e}n$ over $R.$\\

\noindent\textbf{Theorem 4.3.} \emph{Let $\alpha \in
\mathbb{F}_{p^{2m}}$ such that $\alpha\bar{\alpha}=1$. Let
$x^{n}-\beta=\prod_{i=1}^{r}f_{i}(x)$ be the unique factorization of
$x^{n}-\beta$ into a product of monic basic irreducible pairwise
coprime polynomials in $\mathbb{F}_{p^{2m}}[x],$ where
$\emph{gcd}(n,p)=1$. If $C$ is an $\alpha(1+u)$-constacyclic code of
length $N=p^{e}n$ over $R$, then
$C=\langle\prod_{i=1}^{r}f_{i}^{k_{i}}(x)\rangle$, where $0\leq
k_{i}\leq p^{e+1}$.  Then $C^{\perp
H}=\langle\prod_{i=1}^{r}[f_{i}^{\dagger}(x)]^{p^{e+1}-k_{i}}\rangle$
and $|C^{\perp H}| = p^{2mt}$, where $t = \sum\nolimits_{i=1}^{r}{
k_{i}\deg (f_i(x))}. $}\\

\noindent\textbf{Proof.}  Let
$G=\langle\prod_{i=1}^{r}[f_{i}^{\dagger}(x)]^{p^{e+1}-k_{i}}\rangle$
be an
 ideal of $R[x]/\langle x^{N}-\alpha(1+u)\rangle $. By direct computation, we get
\[
\prod_{i=1}^{r}f_{i}^{k_{i}}(x)\cdot\left(\prod_{i=1}^{r}
[f_{i}^{\dagger}(x)]^{p^{e+1}-k_{i}}\right)^{\dagger}=\prod_{i=1}^{r}f_{i}^{p^{e+1}}(x)=[x^{p^{e}n}-\beta]^{2}=0.
\]

\noindent Hence, $G\subseteq C^{\perp}$.  For each $i$, let $a_{i} $
denote the constant of $f_{i}( x)$. Since $\prod_{i=1}^{r}f_{i}(x)=
x^n - \beta$, we have $\prod_{i=1}^{r}a_{i}= \beta$. It follows that
each ${a_{i}}$ is an invertible element of $R$ and ${\bar{a}_{i}}$
is a leading coefficient of ${f_{i}^{\dagger}(x)}$. Hence, for each
$i$, there exists a suitable invertible element $b_{i}$ of $R$ such
that $b_{i} f_{i}^{\dagger}(x)$ is a monic polynomial. Note that
$b_{i} = \bar{a}_{i}^{-1}$ and $\prod_{i=1}^{r}b_{i} =
\prod_{i=1}^{r}\bar{a}_{i}^{-1}=\beta$. So

\begin{eqnarray*}
\prod\nolimits_{i=1}^{r}b_{i} f_{i}^{\dagger}(x)&=&
\prod\nolimits_{i=1}^{r}b_{i}
\cdot\prod\nolimits_{i=1}^{r}\overline{f_{i}^{*}(x)}\\
&=& \beta x^{n}\prod\nolimits_{i=1}^{r}\overline{f_{i}(x^{-1})}\\
&=& \beta x^n(\overline{{x^{ - n} - \beta}})\\ &=&x^n - \beta.
\end{eqnarray*}

\noindent Thus, $b_{i} f_{i}^{\dagger}(x)$ ($1\leq i\leq r$) are
monic basic irreducible divisors of $x^n - \beta$ in $R[x]$. Note
that
$$G=\left\langle\prod_{i=1}^{r}[f_{i}^{\dagger}(x)]^{p^{e+1}-k_{i}}\right\rangle=\left\langle\prod_{i=1}^{r}[b_{i}f_{i}^{\dagger}(x)]^{p^{e+1}-k_{i}}\right\rangle,$$

\noindent On the other hand, by Theorem 4.1, we have

\begin{eqnarray*}
|G|&=& p^{2m(2N-\sum_{i=1}^{r}(p^{e+1}-k_{i})\textrm{deg}(b_{i}f_{i}^{\dagger}(x))}\\
&=& p^{2m(2N-\sum_{i=1}^{r}(p^{e+1}-k_{i})\textrm{deg}(f_{i}(x))}\\
&=& p^{2m(\sum_{i=1}^{r}k_{i}\textrm{deg}(f_{i}(x))}\\
&=& p^{2mt}.
\end{eqnarray*}
 Furthermore, from
$|C||G|=p^{4mN}=|C||C^{\bot H}|$, we have that $|G|=|C^{\bot H}|$ . This gives the proof. \qed\\

Next, we consider   the Gray images of $\alpha(1+u)$-constacyclic
codes over $R.$\\

\noindent\textbf{Theorem 4.4.} \emph{Let $C$ be an
$\alpha(1+u)$-constacyclic code of length $N=p^{e}n$ ($n$ prime to
$p$) over $R$. Let $ M=\begin{pmatrix}
a\ & b  \\
s\ & t  \\
\end{pmatrix}\in GL_{2}(\mathbb{F}_{p^{2m}})$,
where $a=t\alpha, s=b\alpha$ and $b\neq t.$ Then $\Phi_{M}(C)$ is an
$\alpha^{2}$-constacyclic code of length $2N$ over
$\mathbb{F}_{p^{2m}}$.}\\

\noindent\textbf{Proof.} Let
$\textbf{c}=(c_{0},c_{1},\ldots,c_{N-1})\in R^N$, where
$c_{i}=r_{i}+uq_{i},$ for $i=0,1,\ldots, N-1.$ From the definition
of Gray map, we have
\begin{eqnarray*}
\Phi_{M}(\textbf{c})&=&(aq_{0}+s(r_{0}+q_{0}),\ldots,aq_{N-1}+s(r_{N-1}+q_{N-1}),bq_{0}+t(r_{0}+q_{0}),\ldots,bq_{N-1}+t(r_{N-1}+q_{N-1})).
\end{eqnarray*}
Therefore
\begin{eqnarray*}
\sigma_{\alpha^{2}}(\Phi_{M}(\textbf{c}))&=&(\alpha^{2}[b
q_{N-1}+t(r_{N-1}+q_{N-1})],
aq_{0}+s(r_{0}+q_{0}),\ldots,aq_{N-1}+s(r_{N-1}+q_{N-1}),
\ldots,\\
&& b q_{0}+t(r_{0}+q_{0}),\ldots,bq_{N-2}+t(r_{N-2}+q_{N-2})).
\end{eqnarray*}
On the other hand
\begin{eqnarray*}
\sigma_{\alpha(1+u)}(\textbf{c})& = &(\alpha(1+u)c_{N-1},c_{0},\ldots,c_{N-2})\\
       & = &(\alpha(r_{N-1}+u(r_{N-1}+q_{N-1})),a_{0}+ub_{0},\ldots,a_{N-2}+ub_{N-2}).
\end{eqnarray*}
We can get
\begin{eqnarray*}
\Phi_{M}(\sigma_{\alpha(1+u)}(\textbf{c}))&=&(\alpha
 a (r_{N-1}+q_{N-1})+\alpha
sq_{N-1},aq_{0}+s(r_{0}+q_{0}),\ldots,aq_{N-2}+s(r_{N-2}+q_{N-2}),\\
&& \alpha b(r_{N-1}+q_{N-1})+\alpha
tq_{N-1}bq_{0}+t(r_{0}+q_{0}),\ldots,bq_{N-2}+t(r_{N-2}+q_{N-2})).
\end{eqnarray*}
Hence, $\sigma_{\alpha^{2}}(\Phi_{M}(\textbf{c}))=
\Phi_{M}(\sigma_{\alpha(1+u)}(\textbf{c}))$. It follows that
$\Phi_{M}(C)$ is an $\alpha^{2}$-constacyclic code of length $2N$.
This completes the proof.
\qed \\

\noindent Consequently, we can obtain the following corollary.\\

\noindent\textbf{Corollary 4.5.} \emph{The Gray image of an
$\alpha(1+ u)$-constacyclic code of length $N=p^{e}n$(n prime to p)
over $R$ under the
Gray map $\Phi_{M}$ is a distance-invariant  $\alpha^{2}$-constacyclic code of length $2N$ over $\mathbb{F}_{p^{2m}}$.}\\

 Define the polynomial Gray map as follows.
\begin{center}
$\Phi_{M}:R[x]/\langle x^{n}-\alpha(1+u)
\rangle\rightarrow\mathbb{F}_{2^{2m}}[x]/\langle x^{2n}-\alpha^{2}
\rangle,$
\end{center}
\begin{center}
$c(x)=r(x)+uq(x)\mapsto(a+s)q(x)+sr(x)+x^{n}[(b+t)q(x)+tr(x)],$
\end{center}
\noindent where $r(x),q(x)\in \mathbb{F}_{2^{2m}}[x]$. It is obvious
that $\Phi_{M}(c(x))$ is the polynomial representation of
$\Phi_{M}(c)$.\\

\noindent\textbf{Theorem 4.6.} \emph{ Let $\alpha \in
\mathbb{F}_{2^{2m}}$ such that $\alpha\bar{\alpha}=1.$ Let $C$ be an
$\alpha(1+u)$-constacyclic code over
$\mathbb{F}_{2^{2m}}+u\mathbb{F}_{2^{2m}}$ of length $N'=2^{e}n$($n$
is odd) and $C=\langle\prod_{i=1}^{r}f_{i}^{k_{i}}(x)\rangle.$ Let $
M=\begin{pmatrix}
t\alpha \ & b  \\
b\alpha \ & t  \\
\end{pmatrix}\in GL_{2}(\mathbb{F}_{p^{2m}})$ and $b\neq t.$
 Then $ \Phi_{M}(C^{\perp H})$ is  an $\alpha^{2}$-constacyclic code over $\mathbb{F}_{2^{2m}}$ of length $2N'$ and  $ \Phi_{M}(C^{\perp H})=\big\langle\prod_{i=1}^{r}[f_{i}^{\dagger}(x)]^{2^{e+1}-k_{i}}\big\rangle.$
 }\\

\noindent\textbf{Proof.}  By Theorem 4.4, we have $C^{\perp
H}=\big\langle\prod_{i=1}^{r}[f_{i}^{\dagger}(x)]^{2^{e+1}-k_{i}}\big\rangle.$
Let $c(x)$ be any element of $C^{\perp H}.$ Then $c(x)$ reduced
modulo $u$ must be in the  $\alpha$-constacyclic code over
$\mathbb{F}_{2^{2m}}$ of length $N$ with $\big\langle
g(x)\big\rangle,$  where
$g(x)=\prod_{i=1}^{r}[f_{i}^{\dagger}(x)]^{h_{i}}$ with $ 1\leq
i\leq r$ and $h_{i}=\textrm{min}\{2^{e+1}-k_{i},2^{e}\}. $ Hence
$c(x)$ can be written as $c(x)=g(x)f_{1}(x)+uf_{2}(x)$, where
$f_{1}(x),f_{2}(x)\in \mathbb{F}_{2^{2m}}[x]$. Then\\

$\Phi_{M}(c(x))=(b+t)\alpha f_{2}(x)+b \alpha
f_{1}(x)g(x)+x^{n}[(b+t)f_{2}(x)+tf_{1}(x)g(x)] $

\qquad\qquad$=(b+t) f_{2}(x)(x^{n}-\alpha)+f_{1}(x)g(x)(b
\alpha+x^{n}t).$

\noindent Since  $g(x)|x^{n}-\alpha$, there exists $h(x)\in
\mathbb{F}_{2^{2m}}[x]$ such that $x^{n}-\alpha=g(x)h(x)$. It
follows that

$\Phi_{M}(c(x))=(b+t) f_{2}(x)(x^{n}-\alpha)+f_{1}(x)g(x)(b
\alpha+x^{n}t)$

\qquad\qquad $=(b+t) f_{2}(x)g(x) h(x)+f_{1}(x)g(x)(b
\alpha+x^{n}t)$

\qquad\qquad $=g(x)[(b+t) f_{2}(x)h(x)+f_{1}(x)(b \alpha+x^{n}t)].$

\noindent So, $\Phi_{M}(C^{\bot H})\subseteq\langle g(x)\rangle$.
Now, if $m(x)\in\Phi_{M}(C^{\bot H})$, then $m(x)=k(x)g(x),$ where
$k(x)\in \mathbb{F}_{2^{2m}}[x]$. Let $g(x)s_{1}(x)=x^{n}-\alpha,$
 for some  $s(x) \in
R[x]$ satisfies $s(x)$ reduced modulo $u$ is $s_{1}(x).$ Thus
$$m(x)s_{1}(x)x^{n}=k(x)x^{n}(x^{n}-\alpha)=\Phi(u k(x)).$$ So,  $uk(x)\in C^{\bot H}$.
Hence $uk(x)=\langle u
\prod_{i=1}^{r}[f_{i}^{\dagger}(x)]^{t_{i}}\rangle,$ where
$t_{i}=2^{e+1}-k_{i}-\textrm{min}\{2^{e+1}-k_{i},2^{e}\}. $ This
shows that $k(x)=\prod_{i=1}^{r}[f_{i}^{\dagger}(x)]^{t_{i}}I(x)$
for some polynomial $I(x)\in \mathbb{F}_{2^{2m}}[x]$. Hence

$$ m(x)= \prod_{i=1}^{r}[f_{i}^{\dagger}(x)]^{t_{i}}I(x)\prod_{i=1}^{r}[f_{i}^{\dagger}(x)]^{h_{i}}=I(x)\prod_{i=1}^{r}[f_{i}^{\dagger}(x)]^{2^{e+1}-k_{i}}.$$

 \noindent  This shows that
\begin{eqnarray*}
\Phi_{M}(C^{\bot H})\subseteq \left\langle
\prod_{i=1}^{r}[f_{i}^{\dagger}(x)]^{2^{e+1}-k_{i}}\right\rangle.
\end{eqnarray*}
Computing their cardinalities, we get that $ \Phi_{M}(C^{\bot H})=
\left\langle \prod_{i=1}^{r}[f_{i}^{\dagger}(x)]^{2^{e+1}-k_{i}}
\right\rangle.$
 \qed \\

Now, we give an example to illustrate the above results as follows.
Comparing with the previously known quaternary linear code in
Ref.[13], our
obtained $4$-ary linear code is optimal.\\
 \noindent\textbf{Example
4.7.} \ \ Let $\mathbb{F}_{4}=\{0,1,\omega,\omega^{2}=1+\omega\}$.
Consider an $\omega(1+u)$-constacyclic code over
$\mathbb{F}_{4}+u\mathbb{F}_{4}$ of length $10$. Note
 \begin{eqnarray*}
 x^{10}-\omega=(f_{0}(x))^{2}(f_{1}(x))^{2}(f_{2}(x))^{2}
 \end{eqnarray*}
 over
$\mathbb{F}_{4}+u\mathbb{F}_{4},$  where $f_{0}(x)=x+\omega,$
$f_{1}(x)=x^2+x+\omega^{2},$ $f_{2}(x)=x^2+\omega^{2}x+\omega^{2}.$
 Let $C=\left\langle f_{0}(x)(f_{1}(x))^{3}(f_{2}(x))^{4} \right\rangle$.  Then $C^{\perp H}=\left\langle (f_{0}^{\dagger}(x))^{3}f_{1}^{\dagger}(x)
 \right\rangle,$ where $f_{0}^{\dagger}(x)=f_{0}(x),\ f_{1}^{\dagger}(x)=f_{1}(x).$
Take $M=\begin{pmatrix}
\omega t\  &  b \\
\omega b\  &  t \\
\end{pmatrix}$, where $ b,t\in \mathbb{F}_{4}^{*}$ and $b\neq t.$
Computation of the Hamming distances of $ \Phi_{M}(C^{\bot H})$ with
the help of the computer algebra system MAGMA, we get the Hamming
distance of $\Phi_{M}(C^{\bot H})$ is $4.$
 Hence, the Gray image $ \Phi_{M}(C^{\bot H})$ is an  $\omega^{2}$-constacyclic code  with $[20, 15, 4]$ over $\mathbb{F}_{4}.$ Compared  with the previously known
 quaternary linear
code in Ref.[13], which is an optimal
 code.\\

Let $\eta$ be a primitive element of $\mathbb{F}_{p^{2m}}$, and let
$\alpha\in \mathbb{F}_{p^{2m}}$ with order $r>1$ and $r\mid(p^m+1)$,
then $\bar{\alpha} \alpha=1$. Since $\gcd(p^e, r )=1$, there is $f$
with $1\leq f\leq p^m$ such that $p^e f \equiv 1 \pmod{r}$. Suppose
$\beta=\alpha^{f}$, then $\beta^{2^e}=\alpha^{p^ef}=\alpha$.
Clearly, $\beta \in \mathbb{F}_{p^{2m}}$ with order $r$. Let
$\delta$ be a primitive $rn$-th root of unity in some extension
field of $\mathbb{F}_{p^{2m}}$ such that $\delta^{n}=\beta$. Then
$$x^{n}-\beta=\prod\limits_{i=0}^{n-1}(x-\delta^{1+ir}).$$ Let
$\Gamma$ be a complete set of $p^{2m}$-cyclotomic coset
representatives modulo $rn$ and $I=\{i \in \Gamma| i\equiv 1\pmod{r}
\}$. For each $i\in I$, let $C_i= \left\{
{i,ip^{2m},\ldots,i(p^{2m})^{\ell_i-1}} \right\}$ be the
$p^{2m}$-cyclotomic coset modulo $r n$ containing $i$, where
$\ell_i$ is the smallest positive integer such that $i\equiv i
(p^{2m})^{\ell_i} \pmod{rn}$. A cyclotomic coset $C_i$ is called
symmetric if $r n- p^{m}i \in C_i$ and asymmetric otherwise.
Asymmetric cosets $C_i$ and $C_{-p^m i}$ come in pair, we use $(C_i,
C_{-p^m i} )$ to denote such an asymmetric pair. Let $M_{i}(x)$
denote the minimal polynomial of $\delta^i$ over
$\mathbb{F}_{p^{2m}}$, then
$$M_i(x)=\prod_{j\in C_i}(x-\delta^j).$$
It is easily verified that there are $I_1, I_2\subset I$ such that
$$x^n-\beta =\prod_{i\in I_1}M_{i}(x) \prod_{i\in I_2} M_{i}(x) M_{-p^m i}(x). $$ By Theorem 4.3, we know that any $\alpha(1+u)$-constacyclic code of
length $N=p^{e}n$ over $R$ has the form $$C=\big\langle\prod_{i\in
I_{1}}M_i^{k_i}(x)[ \prod_{l\in
I_{2}}M_l^{r_l}(x)M_{-p^{m}l}^{s_l}(x)]\big\rangle,$$ where $ 0 \leq k_{i},r_{l},s_{l}\leq p^{e+1}.$ \\

\noindent\textbf{Theorem 4.8.} \emph{Let $\alpha \in
\mathbb{F}_{p^{2m}}$ such that $\alpha\bar{\alpha}=1$ and let
$$C=\big\langle\prod_{i\in
I_{1}}M_i^{k_i}(x)\big[ \prod_{l\in
I_{2}}M_l^{r_l}(x)M_{-p^{m}l}^{s_l}(x)\big]\big\rangle$$ be an
$\alpha(1+u)$-constacyclic code of length $N=p^{e}n$ over $R$, where
$\emph{gcd}(n,p)=1$  and $0\leq k_{i},r_{l},s_{l}\leq p^{e+1}$. Then
$$C^{\bot H}=\big\langle\prod_{i\in
I_{1}}(M_{i}^{\dag}(x))^{p^{e+1}-k_i}[ \prod_{l\in
I_{2}}(M_{l}^{\dag}(x))^{p^{e+1}-r_l}(M_{-p^{m}l}^{\dag}(x))^{p^{e+1}-s_l}]\big\rangle$$
 and $|C^{\perp H}| = p^{2m \varepsilon}$, where  $\varepsilon=\sum_{i\in
I_{1}}k_{i}\emph{deg}[M_{i}(x)]+\sum_{l\in
I_{2}} \{r_{l}\emph{deg}[M_{l}(x)]+s_{l}\emph{deg}[M_{-p^{m}l}(x)]\}.$ }\\

\noindent\textbf{Theorem 4.9.} \emph{ Let $\alpha \in
\mathbb{F}_{p^{2m}}$ such that $\alpha\bar{\alpha}=1$ and let
$$C=\big\langle\prod_{i\in
I_{1}}M_i^{k_i}(x)\big[ \prod_{l\in
I_{2}}M_l^{r_l}(x)M_{-p^{m}l}^{s_l}(x)\big]\big\rangle$$ be an
$\alpha(1+u)$-constacyclic code of length $N=p^{e}n$ over $R$, where
$\emph{gcd}(n,p)=1$  and $0\leq k_{i},r_{l},s_{l}\leq p^{e+1}$.
 Then   $C \subseteq C^{\bot H}$ if and only if \ \
 $p^{e} \leq k_{i}\leq p^{e+1}$ and $p^{e+1} \leq r_l+s_l\leq p^{e+2}.$
 }\\

\noindent\textbf{Proof.} It is known that $$C^{\bot
H}=\big\langle\prod_{i\in I_{1}}(M_{i}^{\dag}(x))^{p^{e+1}-k_i}\big[
\prod_{l\in
I_{2}}(M_{l}^{\dag}(x))^{p^{e+1}-r_l}(M_{-p^{m}l}^{\dag}(x))^{p^{e+1}-s_l}\big]\big\rangle.$$
Therefore,  $C \subseteq C^{\bot H}$ if and only if $$\prod_{i\in
I_{1}}(M_{i}^{\dag}(x))^{p^{e+1}-k_i}\big[ \prod_{l\in
I_{2}}(M_{l}^{\dag}(x))^{p^{e+1}-r_l}(M_{-p^{m}l}^{\dag}(x))^{p^{e+1}-s_l}\big]$$
divides $$\prod_{i\in I_{1}}M_i^{k_i}(x)\big[ \prod_{l\in
I_{2}}M_l^{r_l}(x)M_{-p^{m}l}^{s_l}(x)\big].$$ Comparing the
indexes, we
 obtain that $p^{e} \leq k_{i}\leq p^{e+1}$ and $p^{e+1} \leq r_l+s_l\leq p^{e+2}$. \qed\\

Now, we give an example to illustrate the results of Theorems 4.8 and 4.9 as follows.\\
\noindent\textbf{Example 4.10.} Consider an
$\omega(1+u)$-constacyclic code over
$\mathbb{F}_{4}+u\mathbb{F}_{4}$ of length $34$. The set
$\{0,1,2,3,6\}$ is a complete set of $4$-cyclotomic coset
representatives modulo $17$. In
$(\mathbb{F}_{4}+u\mathbb{F}_{4})[x]$,
$$x^{34}-\omega=(M_{0}(x)M_{1}(x)M_{2}(x)M_{3}(x)M_{6}(x))^{2},$$
where
\begin{eqnarray*}
&&M_{0}(x)=x+\omega, \\
&&M_{1}(x)=x^{4}+(1+u)x^{3}+\omega
x^{2}+(1+u)x+\omega,\\
&&M_{2}(x)=x^{4}+(1+u)x^{3}+\bar{\omega}
x^{2}+(1+u)x+\omega,\\
&&M_{3}(x)=x^{4}+(\omega+u\omega)x^{3}+
x^{2}+(\omega+u\omega)x+\omega,\\
&&M_{6}(x)=x^{4}+(\bar{\omega}+u\bar{\omega})x^{3}+
x^{2}+(\bar{\omega}+u\bar{\omega})x+\omega.
\end{eqnarray*}
Let $C=\langle (M_{0}(x))^{2}(M_{1}(x))^{4}(M_{3}(x))^{4}
(M_{6}(x))^{2}\rangle$. By Theorem 4.8,  the Hermitian dual code of
$C$ is
$$C^{\bot
H}=\langle(M_{0}^{\dag}(x))^{2}(M_{2}^{\dag}(x))^{4}(M_{6}^{\dag}(x))^{2}\rangle,$$
where $M_{0}^{\dag}(x)=M_{0}(x),\ M_{2}^{\dag}(x)=M_{1}(x),\
M_{6}^{\dag}(x)=M_{3}(x).$  By Theorem 4.9, we obtain  $C \subseteq C^{\bot H}.$  \\

We use the notation $[[n,k, d]]_{q}$ to denote a $q$-ary quantum
error-correcting code for $n$ qubits having $q^{k}$ codewords and
the minimum distance $d$. For an $[[n,k,d]]_{q}$ quantum code $C,$
the parameters
 satisfies the Singleton bound, i.e., $k\leq n-2d+2$.
If a quantum code attains this bound, i.e., $k=n-2d+2$, it is called
a
quantum maximum-distance-separable (MDS) code. Now, we recall the Hermitian quantum code construction:\\

\noindent\textbf{Theorem 4.11.} (Hermitian construction [8]) \emph{
If there exists a $\mathbb{F}_{q^{2}}$-linear code $C=[2N',k]_{q^2}$
    such that $C \subseteq C^{\bot_{H}}$, then there exists a $q$-ary $[[2N', 2N'-2k, \geq d^{\bot H}]]_{q}$ quantum code,
    where $d^{\bot H}$ is the minimum Hamming weight of $C^{\bot_{H}}$.}\\

Combining Theorem 4.6 with Theorem 4.10, we see that a Hermitian
 $\alpha^{2}$-constacyclic self-orthogonal code of length $2N'$ over
$\mathbb{F}_{2^{2m}}$ can be constructed from a Hermitian
 $\alpha(1+u)$-constacyclic self-orthogonal code over $R$ of length
$N'$  under the Gray map $\Phi_{M}$. According to Theorem 4.11, we
can obtain the following theorem, which
can be used for construction of quantum codes.\\

\noindent\textbf{Theorem 4.12.} \emph{Let $\alpha \in
\mathbb{F}_{2^{2m}}$ such that $\alpha\bar{\alpha}=1$ and let
$$C=\big\langle\prod_{i\in
I_{1}}M_i^{k_i}(x)\big[ \prod_{l\in
I_{2}}M_l^{r_l}(x)M_{-2^{m}l}^{s_l}(x)\big]\big\rangle$$ be an
$\alpha(1+u)$-constacyclic code of length $N'=2^{e}n$ ($n$ is odd)
over $\mathbb{F}_{2^{2m}}+u\mathbb{F}_{2^{2m}}$, and size $2^{2k},$
where $0\leq k_{i},r_{l},s_{l}\leq 2^{e+1}$.  Let $M=\begin{pmatrix}
t\alpha\  &  b  \\
b\alpha\  &  t \\
 \end{pmatrix} \in GL_{2}(\mathbb{F}_{2^{2m}})$, where $b,t\in
\mathbb{F}_{2^{2m}}^{*}$ and $b\neq t$ . Suppose that $2^{e} \leq
k_{i}\leq 2^{e+1}$ and $2^{e+1} \leq r_l+s_l\leq 2^{e+2}$. Then
there exists a $2^{m}$-ary quantum code with parameters $[[2N',
2N'-2k, \geq d^{\bot H}]]_{2^{m}},$ where $d^{\bot H}$ is the
minimum Hamming distances
of the $\alpha^{2}$-constacyclic code  $\Phi_{M}(C^{\bot H}).$ }\\

Now, we give an example to illustrate the above results as
follows.\\
 \noindent\textbf{Example
4.13.} \ Let $\mathbb{F}_{4^{2}}=\{0,1,\omega,\omega^{2},\ldots,
\omega^{14}\}$ be a finite field with sixteen elements, where
$\omega^{4}=\omega+1.$ Consider a Hermitian
$\omega^{6}(1+u)$-constacyclic self-orthogonal code with length $6$
over $\mathbb{F}_{4^{2}}+u\mathbb{F}_{4^{2}}.$  Let $C$ be an
$\omega^{6}(1+u)$-constacyclic code of length $6$ over
$\mathbb{F}_{4^{2}}+u\mathbb{F}_{4^{2}}.$ Then
$\beta=(\omega^{6})^{2^{4-1}}=\omega^{3}.$
\\
Note that\\
 \begin{eqnarray*}
 x^{6}-\omega^{3}=(f_{0}(x))^{2}(f_{1}(x))^{2}(f_{2}(x))^{2}
\end{eqnarray*}
\\ over $\mathbb{F}_{4^{2}}+u\mathbb{F}_{4^{2}},$ where $f_{0}(x)=x+\omega^{3},$ $f_{1}(x)=x+\omega^{8},$ $f_{2}(x)=x+\omega^{13}.$
Let
$$C=\left\langle (f_{0}(x))^{4}(f_{1}(x))^{4}(f_{2}(x))^{2}\right\rangle.$$ By Theorem 4.9,  $C$  is a Hermitian  $\omega^{6}(1+u)$-constacyclic self-orthogonal code  over
$\mathbb{F}_{4^{2}}+u\mathbb{F}_{4^{2}}$ of length $6.$   Take
$M=\begin{pmatrix}
\omega^{6}t\  &  b \\
\omega^{6}b\  &  t \\
\end{pmatrix}$, where $ b,t\in \mathbb{F}_{4^{2}}^{*}$ and $b\neq t.$
Computation of the Hamming distances of $ \Phi_{M}(C^{\bot H})$ with
the help of the computer algebra system MAGMA, we get the Hamming
distance of $\Phi_{M}(C^{\bot H})$ is $2.$ Using the Hermitian
construction, we can obtain a
 $Q=[[12, 10, 2]]_{4}$ quantum code,  which is  a quantum MDS code.  Thus, our obtained quaternary quantum code is
 with good parameters.

\section{Symplectic construction}
\label{sec:1} In this section, our goal is to construct $p^{m}$-ary
quantum codes with respect to the trace inner product by employing
an $\alpha(1+u)$-constacyclic of length $N=p^{e}n$ ($n$ prime to $p$
) over $R$. In the following, we will
 apply a map $\phi,$  which is  from $R$ to
$\mathbb{F}_{p^{2m}}^{2}.$ The map
$\phi$  is defined as follows.\\

\noindent\textbf {Definition 5.1.} A map $\phi$ on $R$ is defined as
\begin{align*}
\phi: R &\rightarrow \mathbb{F}_{p^{2m}}^{2}\\
 r+u q &\mapsto(q, r),
\end{align*}
where $r, q\in \mathbb{F}_{p^{2m}}$.\\

Let $\mathfrak{A}=(\textbf{a}|\textbf{b})$ be any element in
$\mathbb{F}_{p^{2m}}^{2N},$ where $\textbf{a}=(a_{0}, a_{1}, \cdots,
a_{N-1}),$ $\textbf{b}=(b_{0}, b_{1}, \cdots, b_{N-1})$. Then we
define the symplectic weight of $\mathfrak{A}$ as
$$\textrm{w}_{S}(\mathfrak{A})=|\{i| (a_{i}, b_{i})\neq0,  i=0,1,\ldots, N-1\}|.$$
 For
$\mathfrak{A}=(\textbf{a}|\textbf{b}),\mathfrak{B}=(\textbf{a}'|\textbf{b}')\in\mathbb{F}_{p^{2m}}^{2N}$,
the symplectic distance $d_{S}((\textbf{a}|\textbf{b}),
(\textbf{a}'|\textbf{b}'))$ between two codewords
$(\textbf{a}|\textbf{b})$ and $(\textbf{a}'|\textbf{b}')$ is the
symplectic  weight of the codeword
$(\textbf{a}-\textbf{a}'|\textbf{b}-\textbf{b}')$. The minimum
symplectic distance $d_S(\mathcal{C})$ of $\mathcal{C} $ is defined
as $\min\{w_{S}((\textbf{a}|\textbf{b}),
(\textbf{a}'|\textbf{b}'))|(\textbf{a}|\textbf{b}),(\textbf{a}'|\textbf{b}'))\in
\mathcal{C}, \textbf{a}-\textbf{a}'\neq \textbf{b}-\textbf{b}'\}$.
For any linear code, the minimum symplectic distance
$d_S(\mathcal{C})$ of $\mathcal{C}$ is its minimum symplectic
weight. The  trace inner product of
$\mathfrak{A}=(\textbf{a}|\textbf{b}),\mathfrak{B}=(\textbf{a}'|\textbf{b}')\in\mathbb{F}_{p^{2m}}^{2N}$
is defined to be
$$\langle \mathfrak{A},\mathfrak{B} \rangle_{T}=\textbf{a}\cdot\overline{\textbf{b}'}-\textbf{b}\cdot\overline{\textbf{a}'} \in\mathbb{F}_{p^{m}}.$$
For a $p^{2m}$-ary linear code $\mathcal{C}$ of length $2N$, the
trace dual code $\mathcal{C}^{\perp_{T}}$ is defined as
$$\mathcal{C}^{\perp_{T}}=\{\mathfrak{A}\in\mathbb{F}_{p^{2m}}^{2N}|\langle\mathfrak{A},\mathfrak{B}\rangle_{T}=0\ \textrm{for}\ \textrm{all}\
\mathfrak{B}\in\mathcal{C}\}.$$ A $p^{2m}$-ary linear code
$\mathcal{C}$ of length $2N$ is called trace self-orthogonal if
$\mathcal{C} \subseteq \mathcal{C}^{\bot_{T}}$, and trace self-dual
if
$\mathcal{C}^{\bot_{T}}= \mathcal{C}$. \\

It is known that
$a+ub=0$ if and only if $a=b=0.$ So, from the above definitions, we directly obtain the following  result.\\
\noindent\textbf {Theorem  5.2.} \emph{  Let  $C$ be a linear code
of length $N$ over $R,$  $d_{H}(C)$ be the minimum Hamming  weight
of the linear code $C,$    $d_{S}(\phi(C))$ be the minimum
symplectic weight of the linear code $\phi(C)$ over $
\mathbb{F}_{p^{2m}}.$ Then
$d_{S}(\phi(C))=d_{H}(C).$}\\

\noindent\textbf{Theorem 5.3.} \ \emph{If $C$ is a self-orthogonal
code of length $N$ over $R$ with respect to the Hermitian inner
product, then $\phi(C)$ is a $p^{m}$-ary self-orthogonal code of
length $2N$ with respect to
the symplectic inner product.}\\

\noindent{\it Proof} \ \ Let $\mathfrak{A}=(\textbf{q}|\textbf{r})$
and $\mathfrak{B}=(\textbf{q}'|\textbf{r}')$ be two codewords in
$\phi(C)$. Then there exist $\textbf{c}, \textbf{c}' \in R^{n}$ and
$\textbf{c}=\textbf{r}+u\textbf{q},
\textbf{c}'=\textbf{r}'+u\textbf{q}',$ where
$\textbf{r},\textbf{q},\textbf{r}', \textbf{q}' \in
\mathbb{F}_{p^{2m}}^{N}$. Since $C$ is self-orthogonal with respect
to the Hermitian inner product, it follows that
$$\langle \textbf{c},\textbf{c}'\rangle_{H}  =(\textbf{r}+u\textbf{q})\cdot(\overline{\textbf{r}'}-u\overline{\textbf{q}'})
=\textbf{r}\cdot\overline{\textbf{r}'}-u(\textbf{r}\cdot\overline{\textbf{q}'}-\overline{\textbf{r}'}\cdot\textbf{q})=0.$$
This gives that
$\textbf{r}\cdot\overline{\textbf{q}'}-\overline{\textbf{r}'}\cdot\textbf{q}=0$.
So we get $\langle \mathfrak{A},\mathfrak{B}\rangle_{T}=0$ and
$\mathfrak{B}\in\phi(C)^{\perp_{T}}$. Therefore,
 $\phi(C^{\bot_{H}})\subseteq
\phi(C)^{\bot_{T}}$. If $ C \subseteq C^{\bot_H}$, then $\phi(C)\subseteq  \phi(C^{\bot_{H}})\subseteq \phi(C)^{\bot_{T}}$. The desired result follows. \qed \\

For a code $C$ of length $N$ over $R$, their torsion and residue
codes are codes over $\mathbb{F}_{p^{2m}}$, defined as follows.
$$\textrm{Tor}(C)=\{\textbf{b}\in \mathbb{F}_{p^{2m}}^{N} | u\textbf{b}\in C\},\  \ \ \  \ \textrm{Res}(C) =\{\textbf{a}\in \mathbb{F}_{p^{2m}}^{N}| \exists \textbf{b}\in \mathbb{F}_{p^{2m}}^{n}: \textbf{a}+u\textbf{b}\in C\} .$$
  The reduction modulo
$u$ from $C$ to $\textrm{Res}(C)$  is given by $\varphi :C
\rightarrow \textrm{Res}(C),$  $\varphi(\textbf{a} + u\textbf{b}) =
\textbf{a}.$ It is known that $\varphi$ is well defined and onto,
with $ \textrm{Ker}(\varphi) \cong \textrm{Tor}(C),$ and $\varphi(C)
= \textrm{Res}(C).$ \\

\noindent The code $\textrm{Res}(C)$ is the code over
$\mathbb{F}_{p^{2m}}$ generated by

$$
\left (\begin {array} {ccc} I_{k_{0}} & A & B
\end {array}\right).$$

\noindent The code $\textrm{Tor}(C)$ is the  code over
$\mathbb{F}_{p^{2m}}$ generated by

$$
\left
(\begin {array} {ccc} I_{k_{0}} & A & B \\
0 & I_{k_{1}} & D
\end {array}\right).$$
Note that $C$  has $p^{4m(2k_{0}+k_{1})} $ codewords and that $|
\textrm{Res}(C)|| \textrm{Tor}(C)|=p^{4m(2k_{0}+k_{1})}.$ Hence for
a code $C$ over $R,$  we have $|C| =| \textrm{Res}(C)||
\textrm{Tor}(C)|. $\\

The following Theorem can be found in Ref.[28].

 \noindent\textbf
{Theorem 4.1} \emph{ Let $d_{H}(C)$  be the Hamming distance of the
linear code $C$ over $R$, $d_{H}(\emph{Tor }(C))$ be the Hamming
distance of the code $\emph{Tor }(C).$   Then
$d_{H}(C) = d_{H}(\emph{Tor }(C)) $.}\\

\noindent\textbf {Theorem 5.5.} \emph{ Let $G$ be  a generator
matrix with $ (k_{0}+k_{1})\times N$ for the linear code $C$ over
$R.$ Then $C$ is  a Hermitian self-orthogonal code if and only if
the $ (k_{0}+k_{1})\times (k_{0}+k_{1})$ matrix $G\overline{G}^{T}$
is the zero matrix.}\\

\noindent{\it Proof} \ \ First, suppose that a linear code $C$ has a
generator matrix $G$ with $ (k_{0}+k_{1})\times N $ over $R$ and $C$
is a Hermitian self-orthogonal code over $R$. Then, for any codeword
$\textbf{c} \in C$, since $C$ is a Hermitian self-orthogonal code,
it follows that $\textbf{c} \overline{G}^T=0$. Therefore, $
G\overline{G}^{T}=0$.

Conversely, assume that  $G$ is  a generator matrix with $
(k_{0}+k_{1})\times N $ for the linear code $C$ over $R$ and  $
G\overline{G}^{T}=0.$ For any codewords $\textbf{c}_1\in C $ and
$\textbf{c}_2 \in C$, there are $\textbf{s}_1$ and $\textbf{s}_2\in
R^{k_0+k_1}$ such that $\textbf{c}_1=\textbf{s}_1^T G$ and
$\textbf{c}_2=\textbf{s}_2^T G$. Then $\langle \textbf{c}_1,
\textbf{c}_2  \rangle_H= \textbf{s}_1^T G \overline{G}^T
\overline{\textbf{s}}_2=0$.
  Therefore, $C$ is a Hermitian self-orthogonal code.\qed \\

The following important construction of quantum codes was proposed
by Ashikhmin and Knill in Ref.[1].

\noindent\textbf{Theorem  5.6.} (Symplectic construction [1])
\emph{Let $\mathcal{C}$ be a $q$-ary self-orthogonal $[2N,k]$ code with respect to the trace inner product. Then there exists an $[[N,N-k,d]]_{q}$ quantum code with $d=d_{S}(\mathcal{C}^{\bot H}).$ }\\

Our goal is to construct $p^{m}$-ary self-orthogonal codes with
respect to the trace inner product by employing a
 linear code over
$R$.\\

\noindent\textbf{Theorem 5.7.} \emph{Let $C$  be a linear code over
$R$ of length $N=p^{e}n$ ($n$  prime to $p$ ) with a generator
matrix $G$, type $\{k_{0},k_{1}\}$  and the minimum Hamming weight
of $ C^{\perp H}$ be $d$. If $ G\overline{G}^{T}=0,$  then there
exists a $2^{m}$-ary quantum code with parameters $[[N,
N-2k_{0}-k_{1},
d]]_{p^{m}}$.}\\

In the following, we will apply a Hermitian
$\alpha(1+u)$-constacyclic  self-orthogonal code of length
$N=p^{e}n$ ($n$ prime to $p$ ) over
$R$ to construct a class of $p^{m}$-ary  quantum codes.\\

\noindent\textbf{Theorem 5.8.} \emph{Let $\alpha \in
\mathbb{F}_{p^{2m}}$ such that $\alpha\bar{\alpha}=1$ and let
$$C=\big\langle\prod_{i\in
I_{1}}M_i^{k_i}(x)\big[ \prod_{l\in
I_{2}}M_l^{r_l}(x)M_{-p^{m}l}^{s_l}(x)\big]\big\rangle$$ be an
$\alpha(1+u)$-constacyclic code of length $N=p^{e}n$ over $R$, and
size $p^{2k},$ where $\emph{gcd}(n,p)=1$ and $0\leq
k_{i},r_{l},s_{l}\leq p^{e+1}$. Then $\emph{Tor}(C^{\bot H})$ and
$\emph{Res}(C^{\bot H})$ are both $\alpha$-constacyclic  codes of
length $N$  over $\mathbb{F}_{p^{2m}}$ with the generator of the
following form
$$\emph{Tor}(C^{\bot H})=\big\langle\prod_{i\in I_{1}}(M_{i}^{\dag}(x))^{k_i-p^{e}}\big[
\prod_{l\in I_{2}}(M_{l}^{\dag}(x))^{r_l- \emph{min}\{p^{e}, r_l
\}}(M_{-p^{m}l}^{\dag}(x))^{s_l-\emph{min}\{p^{e},
s_l\}}\big]\big\rangle$$ and
$$\emph{Res}(C^{\bot H})=\big\langle\prod_{i\in
I_{1}}(M_{i}^{\dag}(x))^{p^{e+1}-k_i}\big[ \prod_{l\in
I_{2}}(M_{l}^{\dag}(x))^{p^{e+1}-r_l}(M_{-p^{m}l}^{\dag}(x))^{p^{e+1}-s_l}\big]\big\rangle.$$}\\

\noindent\textbf{Proof.} By Theorem 4.8, we get $$C^{\bot
H}=\big\langle\prod_{i\in I_{1}}(M_{i}^{\dag}(x))^{p^{e+1}-k_i}\big[
\prod_{l\in
I_{2}}(M_{l}^{\dag}(x))^{p^{e+1}-r_l}(M_{-p^{m}l}^{\dag}(x))^{p^{e+1}-s_l}\big]\big\rangle.$$

\noindent By the definition of torsion code, we have
$$\textrm{Tor}(C^{\bot H})=\big\langle\prod_{i\in
I_{1}}(M_{i}^{\dag}(x))^{\textrm{min}\{p^{e+1}-k_i,p^{e+1}\}-\textrm{min}\{p^{e+1}-k_i,p^{e}\}}$$\\
$$\big[ \prod_{l\in
I_{2}}(M_{l}^{\dag}(x))^{\textrm{min}\{p^{e+1}-r_l,p^{e+1}\}-\textrm{min}\{p^{e+1}-r_l,p^{e}\}}(M_{-p^{m}l}^{\dag}(x))^{\textrm{min}\{p^{e+1}-s_l,p^{e+1}\}-\textrm{min}\{p^{e+1}-s_l,p^{e}\}}\big]\big\rangle$$
and  $\textrm{Tor}(C^{\bot H})$ is an $\alpha$-constacyclic  code of
length $N$  over $\mathbb{F}_{p^{2m}}.$ Since  $ 0 \leq
r_{l},s_{l}\leq p^{e+1}$ and $ p^{e} \leq k_{i}\leq p^{e+1},$ it
follows that
$$\textrm{Tor}(C^{\bot H})=\big\langle
\prod_{l\in I_{2}}(M_{l}^{\dag}(x))^{p^{e+1}-r_l-
\textrm{min}\{p^{e}, r_l
\}}(M_{-p^{m}l}^{\dag}(x))^{p^{e+1}-s_l-\textrm{min}\{p^{e},
s_l\}}\big\rangle.$$ Similarly, we can get
$$\textrm{Res}(C^{\bot H})=\big\langle\prod_{i\in
I_{1}}(M_{i}^{\dag}(x))^{p^{e+1}-k_i}\big[ \prod_{l\in
I_{2}}(M_{l}^{\dag}(x))^{p^{e+1}-r_l}(M_{-p^{m}l}^{\dag}(x))^{p^{e+1}-s_l}\big]\big\rangle
$$  and  $\textrm{Res}(C^{\bot H})$
is an   $\alpha$-constacyclic  code of length $N$  over
$\mathbb{F}_{p^{2m}}.$ \qed \\

By Theorems 4.9 and 5.8, we see that a trace self-orthogonal code of
length $2N$ over $\mathbb{F}_{p^{2m}}$ can be constructed from a
Hermitian  $\alpha(1+u)$-constacyclic self-orthogonal code of length
$N$ over $R$ under the
map $\phi$. According to Theorem 5.6, we can obtain a class of $p^{m}$-ary quantum codes.\\

\noindent\textbf{Theorem  5.9.} \emph{   Let $\alpha \in
\mathbb{F}_{p^{2m}}$ such that $\alpha\bar{\alpha}=1$ and let
$$C=\big\langle\prod_{i\in
I_{1}}M_i^{k_i}(x)\big[ \prod_{l\in
I_{2}}M_l^{r_l}(x)M_{-p^{m}l}^{s_l}(x)\big]\big\rangle$$ be an
$\alpha(1+u)$-constacyclic code of length $N=p^{e}n$ over $R$, and
size $p^{2k},$ where $\emph{gcd}(n,p)=1$ and $0\leq
k_{i},r_{l},s_{l}\leq p^{e+1}$.  Suppose that $p^{e} \leq k_{i}\leq
p^{e+1}$ and $p^{e+1} \leq r_l+s_l\leq p^{e+2}$. Then there exists a
$p^{m}$-ary quantum code with parameters $[[N, N-k,  d]]_{p^{m}},$
where $d$ is the
Hamming distance of the code  $\emph{Tor}(C^{\bot H}).$ }\\

Next, let us use some  examples to illustrate our construction
method. We are with the help of the computer algebra system MAGMA to
find   quantum codes  with  good parameters. We construct some
$3$-ary quantum codes by taking advantage of Hermitian
$\omega^{4}(1+u)$-constacyclic  self-orthogonal codes over $\mathbb{F}_{3^{2}}+u\mathbb{F}_{3^{2}}.$\\

\noindent\textbf{Example 5.10.}  Let
$\mathbb{F}_{3^{2}}=\{0,1,\omega,\omega^{2},\ldots, \omega^{7}\}$ be
a finite field with nine  elements, where $\omega^{2}=\omega+1.$
Consider a Hermitian  $\omega^{4}(1+u)$-constacyclic self-orthogonal
code with length $5$ over $\mathbb{F}_{3^2}+u\mathbb{F}_{3^2}.$  Let
$C$ be an $\omega^{4}(1+u)$-constacyclic code of length $5$ over
$\mathbb{F}_{3^{2}}+u\mathbb{F}_{3^{2}}.$ Then
$\beta=(\omega^{4})^{3^{2-1}}=\omega^{4}.$
\\
Note that\\
 \begin{eqnarray*}
 x^{5}-\omega^{4}=f_{0}(x)f_{1}(x)f_{2}(x)
\end{eqnarray*}
\\ over $\mathbb{F}_{3^{2}}+u\mathbb{F}_{3^{2}},$ where $f_{0}(x)=x+1,$ $f_{1}(x)=x^{2}+\omega^{5} x+1,$ $f_{2}(x)=x^{2}+\omega^{7} x+1.$
\\
\noindent Let
$$C=\left\langle (f_{0}(x))^{2}(f_{2}(x))^{2}\right\rangle.$$ By Theorem 4.12, we get $C$  is a Hermitian  $\omega^{4}(1+u)$-constacyclic self-orthogonal code  over
$\mathbb{F}_{3^{2}}+u\mathbb{F}_{3^{2}}$ of length $5$ with
$$C^{\bot H}=\left\langle
(f_{2}(x))^{2} \right\rangle,$$ where  $f_{2}(x)=f_{1}^{\dag}(x).$
By Theorem 5.8, we obtain $ \textrm{Tor}(C^{\bot H})=\big\langle
(f_{2}(x) \big\rangle.$ Computation of the Hamming distances of $
\textrm{Tor}(C^{\bot H})$ with the help of the computer algebra
system MAGMA, we get the Hamming distance of $C^{\bot H}$ is $3.$
Using the Symplectic construction, we can obtain a
 $Q=[[5, 1, 3]]_{3}$ quantum code,  which is  a quantum
MDS code.  Thus, our obtained $3$-ary quantum code is
 optimal.\\

\noindent\textbf{Example 5.11.} Consider a Hermitian
$\omega^{4}(1+u)$-constacyclic self-orthogonal code with length $10$
over $\mathbb{F}_{3^2}+u\mathbb{F}_{3^2}.$  Let $C$ be an
$\omega^{4}(1+u)$-constacyclic code of length $10$ over
$\mathbb{F}_{3^{2}}+u\mathbb{F}_{3^{2}}.$
\\
Note that\\
 \begin{eqnarray*}
 x^{10}-\omega^{4}=f_{0}(x)f_{1}(x)f_{2}(x)f_{3}(x)f_{4}(x)f_{5}(x)
\end{eqnarray*}
\\ over $\mathbb{F}_{3^{2}}+u\mathbb{F}_{3^{2}},$ where $f_{0}(x)=x+\omega^{2},$ $f_{1}(x)=x+\omega^{6} ,$ $f_{2}(x)=x^{2}+\omega x+2,
$ $f_{3}(x)=x^{2}+\omega^{3} x+2,$ $f_{4}(x)=x^{2}+\omega^{5} x+2, $
$f_{5}(x)=x^{2}+\omega^{7} x+2.$
\\
\noindent Let
$$C=\left\langle (f_{0}(x))^{2}(f_{2}(x))^{2}(f_{3}(x))^{2}(f_{4}(x))^{2}\right\rangle.$$ By Theorem 4.12, we get $C$  is a Hermitian  $\omega^{4}(1+u)$-constacyclic self-orthogonal code  over
$\mathbb{F}_{3^{2}}+u\mathbb{F}_{3^{2}}$ of length $10$ with
$$C^{\bot H}=\left\langle
(f_{0}(x))^{2} (f_{4}(x))^{2} \right\rangle,$$ where
$f_{0}(x)=f_{1}^{\dag}(x), f_{4}(x)=f_{5}^{\dag}(x).$ By Theorem
5.8, we obtain $ \textrm{Tor}(C^{\bot H})=\big\langle f_{0}(x)
f_{4}(x) \big\rangle.$ Computation of the Hamming distances of $
\textrm{Tor}(C^{\bot H})$ with the help of the computer algebra
system MAGMA, we get the Hamming distance of $C^{\bot H}$ is $4.$
Using the Symplectic construction, we can obtain a
 $Q=[[10, 4, 4]]_{3}$ quantum code,  which is  a quantum
MDS code.  Thus, our obtained $3$-ary quantum code is
 optimal.\\

We list some quaternary quantum codes which can be constructed
starting from Hermitian $\omega^{4}(1+u)$-constacyclic
 self-orthogonal codes over $\mathbb{F}_{3^{2}}+u\mathbb{F}_{3^{2}}$
in Table I. Compared the parameters of quantum codes available in
Ref.[13], we find that all of our obtained  quantum codes are
optimal.

\begin{table}[htbp]
{\small
\begin{center}
{\small{\bf TABLE I}~~ Some $3$-ary quantum codes from
$\omega^{4}(1+u)$-constacyclic codes over
$\mathbb{F}_{3^{2}}+u\mathbb{F}_{3^{2}}$}
\begin{tabular}{c c c c c c c c c c c c c c c c c c c}
\hline
& Length &   Generator polynomial of $C^{\bot H} $ &    Hamming distance of $C^{\bot H}$   &     Quantum codes   \\
\hline\\
 & 14&     $(x+\omega^{2})^{2}(x^{3}+\omega^{3}x^{2}+\omega^{3}x+\omega^{2})^2$ &   4 &   $[[14,6, 4]]_{3}$\\
  & 20&    $(x+\omega^{3})^2(x^{2}+\omega^{2}x+\omega^{2})^2$&   3 &    $[[20,14, 3]]_{3}$\\
   & 20&    $(x+\omega^{3})^2(x+\omega^{5})^2(x^{2}+\omega^{6}x+\omega^{6})^2$&   4 &    $[[20,12, 4]]_{3}$\\
   & 40&    $(x^{2}+\omega^{5})^{2}(x^{2}+\omega^{3}x+\omega^{7})^{2}(x^{2}+\omega^{5}x+\omega^{3})^{2}(x^{2}+\omega^{7}x+\omega)^{2}$&   6 &   $[[40,24, 6]]_{3}$\\
  & 61 &     $(x^{5}+\omega^{5}x^{3}+\omega^{7}x^{2}+1)^{2}$ &   4 &     $[[61,51, 4]]_{3}$\\
  & 70 &     $(x+\omega^{6})^{2}(x^{2}+\omega^{5}x+2)^{2}(x^{3}+\omega x^{2}+\omega x+\omega^{6})^{2}$ &  4 &     $[[70,58, 4]]_{3}$\\
  & 82 &     $(x+\omega^{6})^{2}(x^{4}+\omega^{7}x^{3}+\omega x+\omega^{3}x+1)^{2}$ &  4 &     $[[82,72, 4]]_{3}$\\
\hline
\end{tabular}
\end{center}}
\end{table}

\noindent\textbf{Code Comparisons 5.12.}  \ \ We obtain the first
three quantum codes $[[14,6, 4]]_{3},$
 $[[20,14, 3]]_{3}$    and
 $[[20,12, 4]]_{3}$  satisfy $2d = n - k. $   Thus, they are all
optimal.   Compared our obtained the other four quantum codes in
Table I with previously known quantum codes in Ref.[13]. We find
that our obtained quantum codes
  $[[40,24, 6]]_{3},$ $[[61,51, 4]]_{3},$ $[[70,58, 4]]_{3},$ and $[[82,72,
  4]]_{3}$ all meet their lower bounds in Ref.[13]. Thus, they are also
  optimal. The example shows that optimal quantum codes also can be constructed from constacyclic codes over finite chain rings.\\

\section{Conclusion}

\label{sec:1} In this paper, we define a class of Gray maps from
$\mathbb{F}_{2^{2m}}+u\mathbb{F}_{2^{2m}}$ to
$\mathbb{F}_{2^{2m}}^2,$ which preserves the Hermitian
self-orthogonal property of linear codes. By employing  Hermitian
construction,  a new class $2^{m}$-ary quantum codes are constructed
from the Hermitian  $\alpha(1+u)$-constacyclic self-orthogonal codes
over $\mathbb{F}_{2^{2m}}+{u}\mathbb{F}_{2^{2m}}.$   We also define
another class of  maps from
$\mathbb{F}_{p^{2m}}+u\mathbb{F}_{p^{2m}}$ to
$\mathbb{F}_{p^{2m}}^2,$ which changes the Hermitian self-orthogonal
property of linear codes over $R$ into the  trace self-orthogonal
property of linear codes over $\mathbb{F}_{p^{2m}}$. By using the
Symplectic construction,   a new class of $p^{m}$-ary quantum codes
are constructed from the Hermitian $\alpha(1+u)$-constacyclic
self-orthogonal codes over
$\mathbb{F}_{p^{2m}}+{u}\mathbb{F}_{p^{2m}}.$  Our research showed
that quantum codes with good parameters also can be constructed from
the Hermitian  $\alpha(1+u)$-constacyclic self-orthogonal codes over
$R$. Our results also enrich the variety of quantum error-correcting
codes. It would be interesting to encourage more researchers to find
more quantum codes with good parameters from constacyclic code over
finite chain rings and to apply them in practice.

\dse{~~Acknowledgement} The authors would like to thank Doctor
 Sun Zhonghua  who gave many helpful suggestions and comments to
greatly improve the presentation of the paper.The authors also wish
to thank the  editor and the anonymous referee whose comments have
greatly improved this paper.

\dse{~~Availability of data and materials} Data sharing not
applicable to this article as no datasets were generated or analysed
during the current study.

\dse{~~Declaration of competing interest} The authors declare that
we have no known competing financial interests or personal
relationships that could have appeared to influence the work
reported in this paper.


%
%



\end{document}